\documentclass[twocolumn,showpacs,preprintnumbers,amsmath,amssymb,superscriptaddress,prl]{revtex4}

\usepackage{graphicx}% Include figure files
\usepackage{dcolumn}% Align table columns on decimal point
\usepackage{bm}% bold math

\begin{document}

\bibliographystyle{apsrev}

\title{Antinodal quasiparticles below and above $T_{\mathrm{Curie}}$ in the CMR oxide La$_{2-2x}$Sr$_{1+2x}$Mn$_2$O$_7$ with $x=0.36$}

\author{S. de Jong}
\email{sdejong@science.uva.nl}
\affiliation{Van der Waals-Zeeman Institute, University of Amsterdam, NL-1018XE
Amsterdam, The Netherlands}
\author{Y. Huang}
\affiliation{Van der Waals-Zeeman Institute, University of Amsterdam, NL-1018XE Amsterdam, The Netherlands}
\author{I. Santoso}
\affiliation{Van der Waals-Zeeman Institute, University of Amsterdam, NL-1018XE Amsterdam, The Netherlands}
\author{F. Massee}
\affiliation{Van der Waals-Zeeman Institute, University of Amsterdam, NL-1018XE Amsterdam, The Netherlands}
\author{W.K. Siu}
\affiliation{Van der Waals-Zeeman Institute, University of Amsterdam, NL-1018XE Amsterdam, The Netherlands}
\author{A. Mans}
\affiliation{Van der Waals-Zeeman Institute, University of Amsterdam, NL-1018XE Amsterdam, The Netherlands}
\author{R. Follath}
\affiliation{BESSY GmbH, Albert-Einstein-Strasse 15, 12489 Berlin, Germany}
\author{O. Schwarzkopf}
\affiliation{BESSY GmbH, Albert-Einstein-Strasse 15, 12489 Berlin, Germany}
\author{M. S. Golden}
\email{mgolden@science.uva.nl}
\affiliation{Van der Waals-Zeeman Institute, University of Amsterdam, NL-1018XE
Amsterdam, The Netherlands}

\date{\today}

\begin{abstract}
In light of recent conflicting angle resolved photoemission
studies on the bilayered colossal magnetoresistant (CMR) manganite
La$_{2-2x}$Sr$_{1+2x}$Mn$_2$O$_7$ ($0.36\leq x\leq0.40$), new
ARPES data are presented for $x=0.36$ and 0.40, showing only for
the former clear quasiparticle-like features at and around the
$(\pi, 0)$--point in $k$--space. The electronic states are clearly
renormalised, both as regards their dispersion relation and
lifetime due to coupling to bosonic degrees of freedom.
Importantly, both the existence of quasiparticles and their
renormalisation survive well into the paramagnetic state, up to 50
K above $T_\mathrm{Curie}$. This argues against strong coupling to
spin modes and raises questions regarding the nature of the
paramagnetic insulating phase.
\end{abstract}

\pacs{74.25.Jb, 75.47.Lx, 79.60.-i}%Electronic structure, Manganites, Photoemission

\maketitle

%introduction:
Recently, there has been an upsurge in interest in bilayer manganates (La$_{2-2x}$Sr$_{1+2x}$Mn$_2$O$_7$) with
$x\approx0.3$-$0.40$ (abbreviated forthwith LSMO). New experimental data has appeared from STM \cite{Renner} and angle
resolved photoemission (ARPES) studies \cite{Shen,Dessau}. In the new photoemission data, `quasi-particle-like peaks'
(QP peaks) have been reported at the Fermi level ($E_{F}$) for doping levels between $0.40$ and $0.36$. The presence
and behavior of QP features are an important clue for resolving the nature of the ferromagnetic metal to paramagnetic
insulator transition and the related, but poorly understood CMR effect that are displayed by these compounds.

In one of the published ARPES datasets QP peaks are observed for $x=0.40$, but only around the Brillouin zone diagonal
(or `nodal' region), while the $(\pi, 0)$ (or `antinodal') region of the Brillouin zone was peak-less and gapped at
$E_{F}$ \cite{Shen}. These observations have given rise to the hypothesis that the LSMO family are `nodal metals', in
analogy to the high $T_{c}$ superconductors (HTSC), implying that the (in)famous pseudo-gap phase is a state of matter
that is not uniquely reserved for the HTSC \cite{Shen}. In this context, the bilayer manganates have even been
connected recently with `latent d-wave superconductivity' \cite{baskaran}.

In clear contradiction, other published data do not exhibit any QP features at all for $x=0.40$
\cite{Dessau,Dessau_science}. In addition, the existence of QP peaks was reported in an entirely different region of
the Brillouin zone for the doping levels $x=0.36$ and $x=0.38$, namely around the $(\pi, 0)$ point, which is thus
non-gapped \cite{Dessau}.

Can these seemingly contradicting results be reconciled with one another? One suggestion \cite{Dessau} has been
connected with the canting of the magnetic moments on the Mn sites on going from $x=0.40$ to $x=0.36$, resulting in
increased three dimensionality for the latter. Indeed, a subtle interplay between orbital and electronic degrees of
freedom causes LSMO to have a complex (magnetic) phase diagram as a function of doping \cite{ling,kimura}. However, at
present it remains unclear as to whether the \emph{gradual} structural changes observed on going from $x=0.40$ to
$0.36$ are really able to quantitatively  explain the seemingly diametrically opposing paradigms of `nodal-metal'
versus an `anti-nodal metal' for the same group of materials.

Important and far reaching statements have been made about the electronic nature of LSMO in connection to the physics
of the HTSC, based on a nodal metal picture that has proven not to be generic for the entire range of doped bilayer
manganates. Thus, in order to get clarity about the validity of the pictures presented, it is evident that additional
high quality experimental data is required. In this letter we present such measurements, carried out on LSMO with
$x=0.36$ and $0.40$. We show clearly the existence of (renormalised) QP features in the $(\pi, 0)$ region of $k$-space
for $x=0.36$ and the absence of such features in the same k-space region for $x=0.40$. Furthermore, for the first time,
we explore the temperature dependence of these antinodal QPs.

\begin{figure} [!tb]
\begin{center}
\includegraphics[width=1.0\columnwidth]{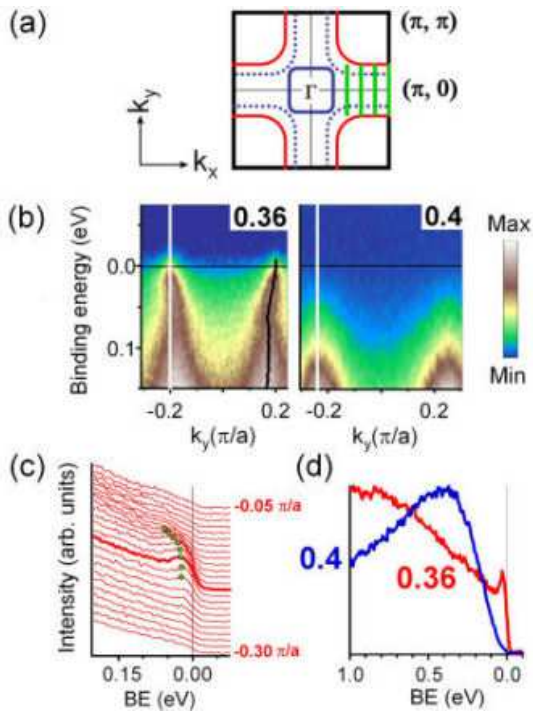}
\caption{\label{fig:EDM} (\textbf{a}) Brillouin zone sketch
showing the bonding (red) and anti--bonding (AB) band (blue ) and
the $3d_\mathrm{{z^2-r^2}}$ electron pocket (blue solid line). The
$k$-space cuts are shown in green. (\textbf{b}) ARPES spectra for
LSMO $x=0.36$ (\emph{h$\nu=56$}~eV) and LSMO $x=0.40$
(\emph{h$\nu=80$}~eV) taken at $k_{x}=\pi/a$, $T=25$~K. For
$x=0.36$: a black curve tracks the AB band MDC maxima. Note the
absence of spectral weight at $E_{F}$ for $x=0.40$. (\textbf{c})
Stack-plot of EDCs for $x=0.36$ ($k_{F}$-EDC is bold), showing a
small dispersive peak close to $E_{F}$. (\textbf{d}) comparison of
EDC's at $k_{F}$ [indicated by the white lines in panel (b)] for
$x=0.36$ (red) and 0.40 (blue) [\onlinecite{footnote_kF}]. }
\end{center}
\end{figure}

%experimental:
Experiments were performed at the U125-PGM1 beamline at BESSY, using an SES100 analyzer. The experimental resolution
was 30~meV in energy and 0.01~$\pi/a$ in $k$ at the excitation energies used. High quality single crystals of LSMO were
investigated, grown in Amsterdam using the optical floating zone technique. Both doping levels studied are paramagnetic
(PM) insulators at high temperatures, but become (poor) ferromagnetic metals below the Curie temperature $T_{C}$, which
was determined using SQUID magnetometry to be 131~K for $x=0.36$ and 125~K for $x=0.40$. The crystals were cleaved at
$T<40$~K, and the base pressure was lower than $1\times10^{-10}$~mbar. Very sharp, tetragonal low energy electron
diffraction patterns were obtained from all the measured cleavage surfaces.

%results and discussion:
A sketch of the Brillouin zone of LSMO 0.36 is depicted in
Fig.~\ref{fig:EDM}(a) \cite{huang,dessau2}. In
Fig.~\ref{fig:EDM}(b) a typical energy distribution map (EDM) is
shown as obtained from a zone face cut with $k_{x}=\pi/a$ for both
LSMO $x=0.36$ and $0.40$. The former displays a dispersive feature
[the Mn $3d_\mathrm{{x^2-y^2}}$ anti--bonding (AB) band] crossing
$E_{F}$ at $k_{y}\approx\pm0.2~\pi/a$. A stack plot of the
relevant energy distribution curves (EDCs) forms panel (c) of Fig.
1, showing a small peak at low binding energy (BE) dispersing
towards $E_{F}$ on top of a background that is increasing to
higher BE. Analogously to Refs. [\onlinecite{Shen},
\onlinecite{Dessau}], we refer hereafter to the small peak in the
EDCs as a `quasi-particle peak', without further discussion, for
example, of its width as a function of energy.

The data from LSMO with $x=0.40$ in Fig.~\ref{fig:EDM}(b) show strongly suppressed spectral weight near $E_{F}$, when
compared to the $x=0.36$ data from the same $k$-space location. Although we carried out extensive measurements on the
$x=0.40$ doping level utilising different excitation energies and measurement geometries, we never observed a QP peaks
such as those seen for $x=0.36$. Panel (d) of Fig.~\ref{fig:EDM} brings the differences between the data for $x=0.36$
and 0.40 sharply into focus. We show $k_{F}$-EDCs for both doping levels \cite{footnote_kF}. The QP peak for $x=0.36$
is preceded by a broad hump at higher BE, which has its maximum at around 1~eV. This hump has been connected with the
incoherent part of the single particle spectral function, resulting from the coupling of bosonic modes to the electron
system \cite{Dessau}. For $x=0.40$, the small QP peak is clearly absent and the spectral weight at $E_{F}$, though
non-zero, is very small.

\begin{figure} [!b]
\begin{center}
\includegraphics[width=1.0\columnwidth]{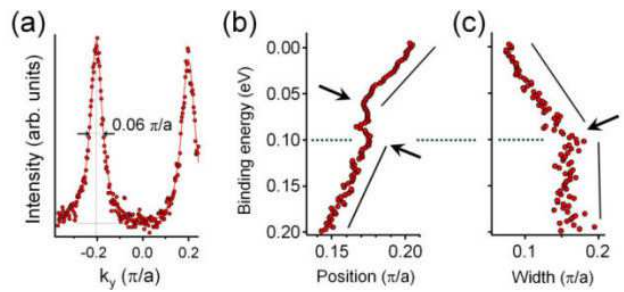}
\caption{\label{fig:MDC} MDC analysis of the $x=0.36$ EDM of
Fig.~\ref{fig:EDM}b. (\textbf{a}) MDC at E$_{\mathrm{F}}$, fitted
with two Lorentzians and a linear background. (\textbf{b})
position of MDC peaks vs.\ BE, clearly exhibiting a
renormalisation in the dispersion at $100$~meV and $50$~meV
(marked with arrows). (\textbf{c}) MDC peak width vs. BE, also
showing a `kink' (arrow) around $\mathrm{BE}=100$~meV.}
\end{center}
\end{figure}

The dispersion of a band can be traced via analysis of the momentum distribution curves (MDCs: horizontal cuts through
an EDM). Such an MDC taken at $E_{F}$ from the $x=0.36$ data of Fig.~\ref{fig:EDM}(b) is depicted in
Fig.~\ref{fig:MDC}(a). Two well-defined peaks are seen with a width of 0.06~$\pi/a$, which corresponds to a mean free
path length of about 35~\AA, would one assume that the width is entirely due to life-time broadening. It is clear from
Fig.~\ref{fig:MDC}(b), which shows the $k_{y}$ positions of the MDC maxima plotted as a function of BE, that the MDC
maxima trace a path deviating significantly from the parabola expected for a non-interacting (`bare') electron band. In
fact, a two-branch dispersion is seen, whereby below 100~meV the velocity alters with respect to the high BE value, and
under 50~meV a new, lower velocity is fully established. Such behaviour in the dispersion is a signal for significant
structure in the real part of the self energy, due to coupling to one or more bosonic modes \cite{kink}.

Within a simple analysis, the dimensionless coupling constant for the interaction with the mode(s) $\lambda$, follows
from $1+\lambda=m_{\mathrm{fast}}/m_{\mathrm{slow}}$. This yields a coupling strength $\lambda$ of the order of unity,
which agrees for the value reported for $x=0.38$ in Ref.\ [\onlinecite{Dessau}]. The fact that the well-defined MDC
feature can be followed over more than 300~meV in BE means that the accepted MDC method - appropriate for a non-gapped
region of $k$-space - can and should be used to estimate $\lambda$ \cite{Bogdanov}. If one were to extract the low BE
velocity from the EDC peak maxima, an artificially large value of $\lambda$ would be arrived at, of order of that
quoted in Ref.\ [\onlinecite{Shen}].

\begin{figure}[!t]
\begin{center}
\includegraphics[width=1.0\columnwidth]{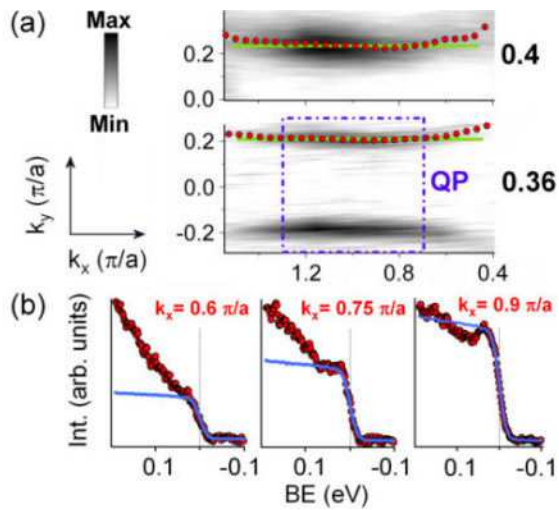}
\caption{\label{fig:Fsmap} (\textbf{a}) Constant BE maps recorded
at $T=25$~K for LSMO $x=0.40$ ($\mathrm{BE}=60\pm15$~meV,
$h\nu=67$~eV) and $0.36$ ($E_{F}\pm10$~meV, $h\nu=56$~eV),
constructed from cuts as indicated in Fig~\ref{fig:EDM}(a). Red
dots indicate $k$-values of the AB bands for the relevant energy
surfaces. The straight green line is a guide to the eye. The
$k$-space region supporting QP peaks in the EDCs is highlighted
using a blue dotted box. (\textbf{b}) $k_{F}$-EDCs for $x=0.36$
for $k_{x}$ as marked. The blue solid line shows the Fermi level
position from a gold reference (normalized to the Fermi edge step
heights).}
\end{center}
\end{figure}

Not only the position, but also the widths of the MDC peaks show a clear `kink' as a function of BE. As can be seen in
Fig.~\ref{fig:MDC}(c), below 100~meV there is a rapid decrease in the MDC width, which reflects a decrease in the
scattering rate. This in turn is evidence of significant structure for this energy and $k$-location in the imaginary
part of the self energy, in keeping with the anomaly in the dispersion discussed above.

By fitting a parabolic dispersion through the peaks of the MDCs at $k_{F}$ and high BE (i.e. excluding the MDCs at
energies around the kink), the bottom of the AB band can be placed at about 400~meV BE. On taking an excitation energy
of 74~eV, it was also possible to observe a second dispersing feature for LSMO $x=0.36$, similar to the one shown in
Fig.~\ref{fig:EDM}(b), but with $k_{F}$ $\approx 0.28~\pi/a$ and a band bottom at around 700-800~meV BE. This second
band can be attributed to the Mn 3d$_\mathrm{{x^2-y^2}}$-based bonding band (BB), in keeping with Ref.\
[\onlinecite{Dessau}]. With this in mind, one can certainly ascribe --at least-- part of the observed high BE spectral
weight for LSMO $x=0.36$ (the `hump') in Fig.~\ref{fig:EDM}(d) to the bonding band.

Up until now we have concentrated on the antinodal or zone face states with $k_{x}=1~\pi/a$. Now in
Fig.~\ref{fig:Fsmap}(a) we show a constant energy map for $E=E_{\mathrm{F}}$ for LSMO $x=0.36$ covering the ($\pi$,0)
point, but also stretching quite some way into the first and second Brillouin zones. The two Fermi surface segments
(both AB band), are nested over a significant range of $0.35~\pi/a$ either side of $k_{x}=\pi/a$, before they start to
deviate significantly from being parallel.

In contrast, the QP peak in the EDCs is resolvable only over a relatively short $k$-range. $k_{F}$-EDCs are shown in
Fig.~\ref{fig:Fsmap}(b), from which one can see that while at $k_{x}=0.75~\pi/a$ there is still a small peak visible on
top of the large incoherent hump, by $k_{x}=0.6~\pi/a$ no peak is observed in the EDCs. Interestingly, despite this,
the characteristic two-branch band dispersion is still present at this point in $k$-space (not shown). We note here
that the disappearance of the QP peak spectral weight at this $k$-value is not associated with the opening of a gap, as
can be seen in Fig.~\ref{fig:Fsmap}(b) from a comparison of the EDCs with the Fermi edge of a polycrystalline gold
film. In addition, an analysis of the $k_{F}$ EDCs by means of symmetrization \cite{symmetrize} indicates no opening of
a gap. For $k$-values closer to $\Gamma$ than $0.6~\pi/a$, the statistics of the data are insufficient to make a
conclusion regarding the opening of a gap. Thus, the data presented here for $x=0.36$ show that at least two thirds in
length of the side of the $(\pi, \pi)$-centered Fermi surface barrels is \emph{ungapped}.

For comparison, we show in Fig.~\ref{fig:Fsmap}(a) a constant energy map for LSMO $x=0.40$, in this case for only one
side of the $\Gamma$-[$\pi$,0] line. Because of the very low spectral weight at E$_{\mathrm{F}}$ for $x=0.40$, the map
shows data integrated over 60~meV below the Fermi level, which results in a broader band. Although for $x=0.40$ no QP
peaks were found, the map resembles that of LSMO $x=0.36$ closely; both show a similar shift in $k_{F}$ as a function
of $k_{x}$ and a relative decrease of spectral weight away from the $(\pi,0)$ point.

\begin{figure} [!tb]
\begin{center}
\includegraphics[width=1.0\columnwidth]{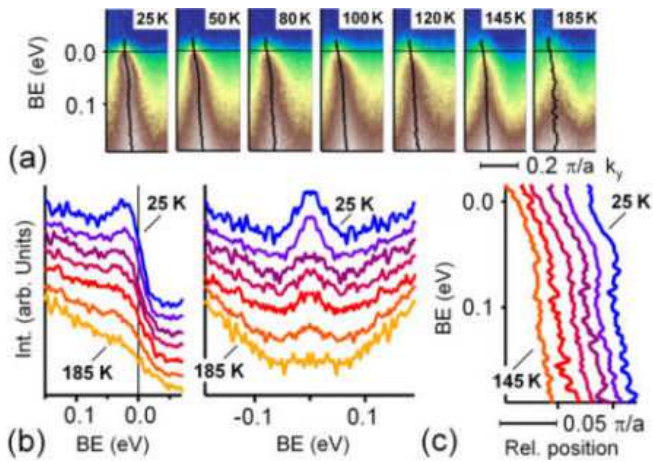}
\caption{\label{fig:Tdependence}(\textbf{a}) T-dependence for LSMO
$x=0.36$ at ($\pi$,0) (\emph{h$\nu$=$56$}~eV, left hand branch).
The MDC peak maxima are overlaid using a black solid line, and are
shown separately (offset) in panel (\textbf{c}). (\textbf{b})
Left: EDCs at $k_{F}$ for all temperatures and right: their
symmetrized versions (all curves offset vertically). }
\end{center}
\end{figure}

Now moving beyond the question as to the applicability of the
epithet 'nodal metal', for these systems, we present the first
data concerning the temperature dependence of the quasiparticle
excitations at the zone face of a bilayer manganate ($x=0.36$).
Fig.~\ref{fig:Tdependence} shows data from the $(\pi,0)$-point
taken at temperatures from 25~K up to 185~K, the latter well above
the $T_{C}$ of 130~K. Panel (a) contains EDMs showing one of the
branches of the AB band. It is clear that -- apart from thermal
broadening -- all the EDMs look very similar. The spectral weight
of the QP peak does decrease steadily with temperature, as can be
seen from the EDCs shown in Fig.~\ref{fig:Tdependence}(b).
Nevertheless, even at 185~K some QP spectral weight still remains,
causing a plateau in the slope of the incoherent hump seen in the
EDC. None of the spectra in Fig.~\ref{fig:Tdependence} show a gap,
as can be seen from the symmetrized $k_{F}$-EDCs displayed in
panel (b). The fact that QP spectral weight near $(\pi,0)$ exists
even 50~K into the paramagnetic region of the phase diagram seems
at odds with its insulating nature, and deserves further study,
both experimentally and theoretically. Additionally, in
Fig.~\ref{fig:Tdependence}(c), one sees that the dispersion of the
band is essentially independent of the temperature: even at 145~K
the effective masses of two dispersion branches remain essentially
unaltered.

These data put important constraints on the identity of the boson(s) coupling to the fermionic system in this region of
$k$-space. The two-branch behaviour and lifetime renormalisation effects are caused by moderately strong
$(\lambda\approx1)$ coupling to largely temperature \emph{independent} collective degrees of freedom. Dealing first
with the spin channel, we mention that fluctuations of the ferromagnetic order parameter would be expected to play an
increasing role close to $T_{C}$: not a trend seen in the two-branch nature of the dispersion data. Magnons can also be
ruled out as the culprit: firstly, the coupling for a half-metal with large exchange interaction would be expected to
be small \cite{SchaeferPRL}, and secondly the magnon energies in LSMO \cite{HirotaPRB} are much lower than the energy
scale of the renormalisation seen in the ARPES data. Long-lived antiferromagnetic clusters have been shown to exist in
LSMO (x=0.40) in the PM phase \cite{PerringPRL}, but they disappear quickly for $T<T_C$, thus excluding one of the
bosonic channels most discussed in the context of the HTSC cuprates. This leaves two possibilities: phonons or orbital
degrees of freedom. The former (in particular bond-stretching modes) have been suggested as the cause in Ref.\
[\onlinecite{Dessau}]. The latter -- certainly one of the leitmotifs in the physics of the manganese oxides -- have
been claimed to lie in an energy range above 100~meV from Raman data \cite{SaitohNature}, although the interpretation
of the Raman data in terms of orbital waves has been strongly contested from optical conductivity data
\cite{GrueningerNature}. From theory \cite{vandenBrinkPRL}, the elementary excitations are expected to be of mixed
phonon/orbiton character, thus if the Raman features in Ref.\ \cite{SaitohNature} at around 150~meV are orbitally
related, they are orbiton satellites of phonon lines, shifted by 70 meV from the bare phonon lines. Therefore, at this
stage the exclusion of orbital or mixed phonon/orbiton excitations from the arena would seem premature. We note that
the suppression of the QP spectral weight around the node in our data [e.g. in Fig.~\ref{fig:Fsmap}(a)] - which is
where the $d_{\mathrm{z^2-r^2}}$ electron pocket comes closest to the $3d_{\mathrm{x^2-y^2}}$ bands - would not be
inconsistent with a possible role for orbital fluctuations.

In summary, we use ARPES to show unambiguously that QP-like excitations exist at and near the $(\pi,0)$ points in the
bilayer manganate La$_{1.28}$Sr$_{1.72}$Mn$_2$O$_7$. In contrast, our data for La$_{1.2}$Sr$_{1.8}$Mn$_2$O$_7$ is
characterised by a strongly suppressed spectral weight at the Fermi level and a lack of QP excitations at all locations
in k-space measured. In La$_{1.28}$Sr$_{1.72}$Mn$_2$O$_7$, the QP features display clear signs of moderately strong
coupling to bosonic degrees of freedom at an energy scale of the order of 50-100 meV. These renormalisation effects at
the Brillouin zone face are remarkably temperature independent, being still visible some 50~K above $T_{C}$. Either
phonon or orbital excitations are possible candidates for the bosonic degrees of freedom involved, whereas AFM or FM
spin fluctuations would appear to be much less likely candidates. Finally, the `nodal-metal' picture, though possibly
valid for LSMO $x=0.40$, is certainly not generically applicable to the bilayer manganates as a family. Any general
parallel between the manganates and the HTSC cuprates based on this nodal metal scenario \cite{Shen,baskaran} should
thus be reconsidered.

Our thanks to the IFW Dresden group for lending us their spectrometer, and to W. Koops and T. J. Gortenmulder for
expert technical support. This work is funded by the FOM (ILP and SICM) and the EU (I3).


\begin{thebibliography}{}
\bibitem{footnote_kF} for $x=0.40$ the $k_{F}$-EDC is taken as that with the greatest intensity close to E$_{\mathrm{F}}$
\end{thebibliography}
\end{document}